\documentclass[psfig]{mn2e}

\usepackage{times}

\title[Radio flaring behaviour of GRO~J$1655-40$]
{The radio flaring behaviour of GRO~J$1655-40$: an analogy with
extragalactic radio sources?}
\author[J. A. Stevens et al.]
{J. A. Stevens$^{1,2}$, D. C. Hannikainen$^{3,4}$, 
  Kinwah Wu$^{2,5}$, R. W. Hunstead$^{5}$ and \and D. J. McKay$^{6,7}$ \\  
$^{1}$Astronomy Technology Centre, Royal Observatory, Blackford Hill,
Edinburgh, EH9 3HJ\\
$^{2}$Mullard Space Science Laboratory, University College London, 
       Holmbury St.~Mary, Surrey, RH5~6NT\\   
$^{3}$Department of Physics and Astronomy, 
       University of Southampton, Southampton, SO17 1BJ \\ 
$^{4}$Observatory, PO Box 14, 00014 University of Helsinki, Finland\\
$^{5}$School of Physics A28,
       University of Sydney, NSW 2006, Australia \\  
$^{6}$Telescope Technologies Ltd, 1 Morpeth Wharf, Birkenhead, Merseyside CH41 1NQ\\
$^{7}$Australia Telescope National Facility, Locked Bag 194, Narrabri, NSW
2390, Australia 
 }  

\date{Received: }

\begin{document} 
 
\def\Mdot{\hbox{$\dot M$}}
\def\Msun{\hbox{M$_\odot$}}
\def\Rsun{\hbox{$R_\odot$}} 
\outer\def\gtae {$\buildrel {\lower3pt\hbox{$>$}} \over 
       {\lower2pt\hbox{$\sim$}}$}
\outer\def\ltae {$\buildrel {\lower3pt\hbox{$<$}} \over
       {\lower2pt\hbox{$\sim$}}$}
\def\rchi{{${\chi}_{\nu}^{2}$}}   
\def\ddt{{d \over {dt}}} 

\maketitle

\begin{abstract}     
At radio frequencies, the current evidence for the microquasar--quasar
connection is based on imaging observations showing that relativistic
outflows/jets are found in both classes of objects.  Some microquasars also
display superluminal motion, further strengthening the view that
microquasars are in fact Galactic miniatures of quasars.  Here we demonstrate
that this connection can be extended to incorporate timing and spectral
observations.  Our argument is based on the striking similarity found in the
radio flaring behaviour of the Galactic superluminal source GRO~J1655$-$40 and
of extragalactic sources, such as the blazar 3C~273.  We find that the
variability of GRO~J1655$-$40 can be explained within the framework of the
successful generalised shock model for compact extragalactic radio sources in
which the radio emission arises from shocked plasma in relativistic jets.
Specifically, the multifrequency flare amplitudes,
time delays and radio polarization position angle measurements are consistent
with the predictions of the growth stage of this model.
\end{abstract}


\section{Introduction}  

Microquasars, here defined as X-ray binaries containing a black hole accreting
material from a companion star, are generally regarded as the Galactic
counterparts of more powerful extragalactic radio sources, such as blazars
and radio galaxies.
\footnote{The blazar class comprises the brightest and most
variable of the extragalactic radio sources. Accordingly, for the purposes of
this study, most observational data for extragalactic sources that allow
comparison with the microquasars pertain to blazars. However, the `unified
scheme' for radio sources postulates that, to first order, the only difference
between radio galaxies, quasars and blazars is one of orientation (e.g. Barthel
1989) so the same physical model should be applicable to all.}  This implies
that the underlying physical processes that operate in these systems are the
same, and thus are manifested in their radiative properties.  At radio
frequencies, the main argument for microquasars being Galactic miniatures of
quasars is based on imaging observations of relativistic outflows in the
former.  This is further strengthened by the fact that some microquasars also
display superluminal motion analogous to their extragalactic counterparts.  For
recent reviews of the radio properties of microquasars, see e.g.\ Mirabel \&
Rodriguez (1998, 1999) and Fender (2001, 2002).

The two unambiguous Galactic superluminal sources to date are GRS~1915+105 and
GRO~J1655$-$40.  Their radio jets were discovered in the mid-1990s (Hunstead et
al. 1994; Reynolds et al. 1994; Mirabel \&
Rodr{\'\i }guez 1994; Tingay et al.\ 1995; Hjellming \& Rupen 1995).  They
often show radio flares during the X-ray active states, and their radio
emission is polarized.  The nature of GRS~1915+105 has been controversial but
recent observations show that it is likely to be a low mass X-ray binary with a
black-hole candidate and K-giant donor star (Greiner et al. 2001). It is
certain that GRO~J1655$-$40 is a binary system containing an evolved F-type
star and a black hole (Orosz \& Bailyn 1997; Soria et al.\ 1998).

Compact extragalactic radio sources usually show power law spectra at
high-frequencies ($> 90$ GHz).  
The radio emission is of synchrotron origin, is often highly polarized, and the
power law spectrum implies a non-thermal energy distribution for the
relativistic electrons and an optically thin emission region.  The spectra
often have a turn-over at low frequencies that is generally explained as the
consequence of optical-depth effects; the emission is absorbed and the source
becomes opaque below a certain critical frequency.  The time evolution of the
low-frequency (typically less than a few GHz) radio spectra was initially
explained by means of a homogeneous cloud composed of relativistic electrons
and magnetic field that expands adiabatically and emits synchrotron radiation
(Shklovsky 1965; van der Laan 1966; Pauliny-Toth \& Kellerman 1966). However,
it was later found that a more satisfactory explanation is provided by models
invoking shocks propagating along a relativistic jet, and hence accelerating
electrons to relativistic energies (e.g.\ Marscher \& Gear 1985; Hughes, Aller
\& Aller 1985, 1989a, b).
  
Like extragalactic radio sources, the Galactic superluminal source
GRO~J1655$-$40 also shows a power law spectrum at radio frequencies that becomes
inverted during flaring (Hannikainen et al.\ 2000).  To date the time-dependent
radio properties of X-ray binaries have mostly been ascribed to the
expanding-cloud (or synchrotron-bubble) model or variations thereof (Hjellming
\& Johnston 1988; Han \& Hjellming 1992; Ball \& Vlassis 1993) although an
internal shock model has been applied successfully to a radio outburst of
GRS~1915+105 (Kaiser, Sunyaev \& Spruit 2000). Since the 1994 radio outburst of
GRO~J1655$-$40 is not readily accommodated by the expanding-cloud model (see
Hannikainen et al. 2000 and Section 2 below) we propose alternatively that
the processes that generate the radio emission from GRO~J1655$-$40 and from
extragalactic radio sources are the same.  We make use of the shock model of
Marscher \& Gear (1985) to explain the spectral evolution of GRO~J1655$-$40
during the 1994 outburst.  The multifrequency variability is discussed in
Section 2, basic features of the model are described in Section 3, and the
quasar-microquasar connection is discussed in Section 4. 
 
Throughout this paper, spectral index, $\alpha$, is defined as 
$S_{\rm \nu} \propto \nu^{\alpha}$ where $S_{\rm \nu}$ is the flux density at
frequency $\nu$.

\section{Time-dependent properties of GRO J1655$-$40}     

GRO J1655$-$40 flared dramatically in 1994 August, the flux density at 843 MHz
increasing from a few hundred mJy to almost 8~Jy in 12 d (Hannikainen et
al.\ 2000). Very Long Baseline Array (VLBA) images at 1.6 GHz taken at several
epochs during the outburst show well-collimated relativistic jets. The jets are
resolved into several knot-like features which expand outwards from a
stationary compact core (Hjellming \& Rupen 1995).

\begin{figure}
\setlength{\unitlength}{1in}
\begin{picture}(3.25,5.0)
\includegraphics{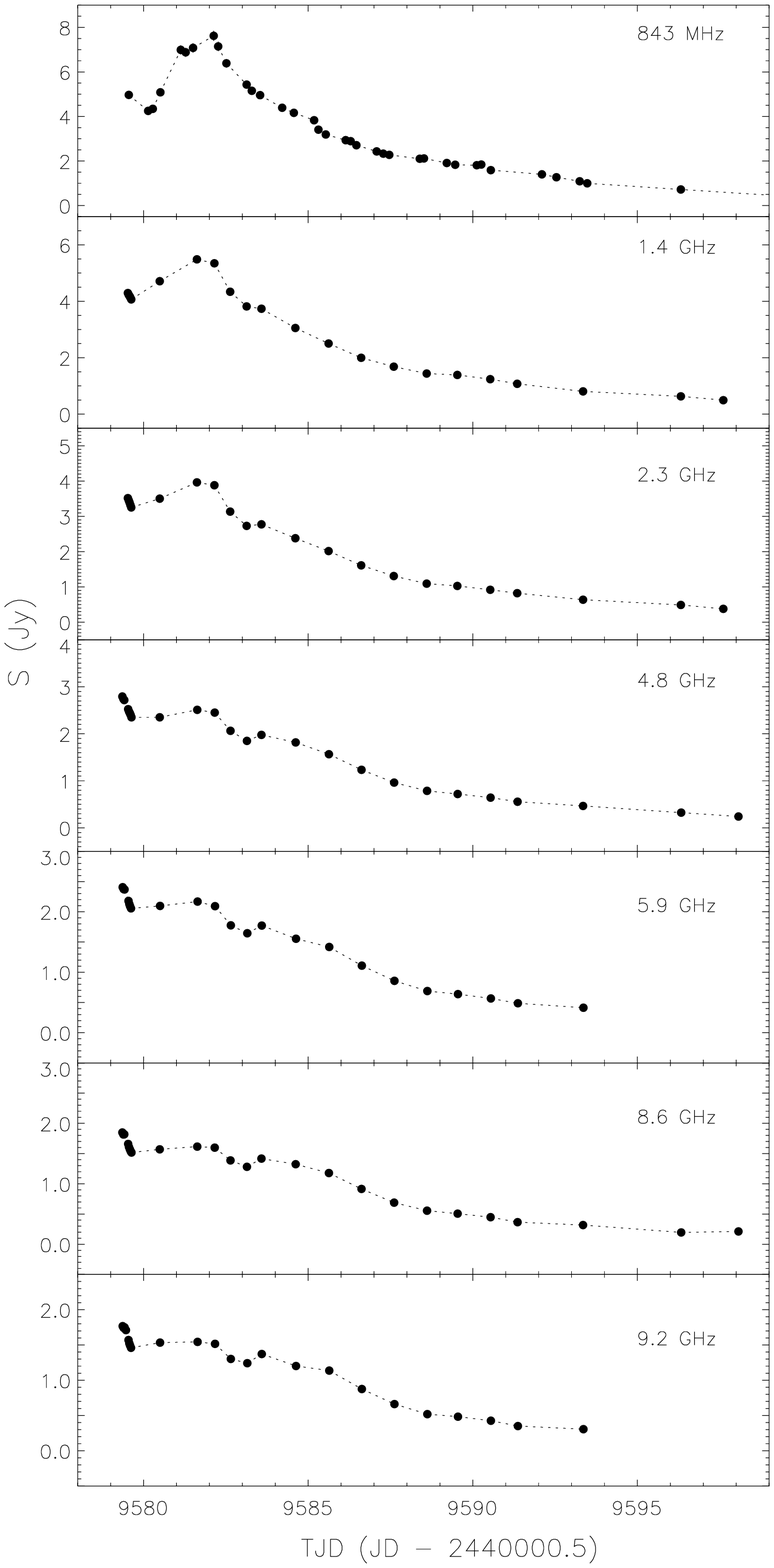}
\end{picture}
\caption{The multifrequency light curves of GRO~J1655$-$40
observed by the Molonglo Observatory Synthesis Telescope (MOST; 843 MHz) and
the ATCA (1.4--9.2 GHz).  Data are adopted from Hannikainen et
al. (2000).}
\label{fig:lc}
\end{figure}

\begin{figure}
\setlength{\unitlength}{1in}
\begin{picture}(3.25,6.0)
\includegraphics{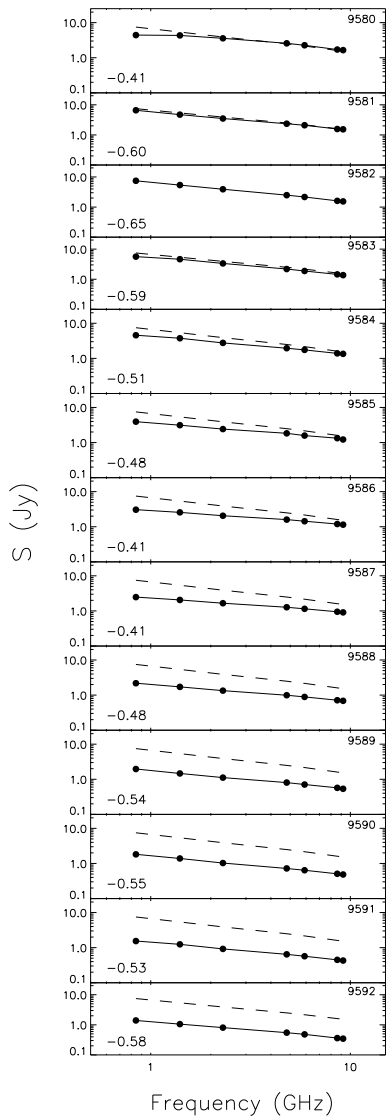}
\end{picture}
\caption{Spectra of GRO~J1655$-$40 (solid lines) are shown at time intervals of 
     one day from TJD 9580 to 9592. The data points are interpolated from the 
     flux density light curves. The spectrum from TJD 9582 (dashed line), the 
     epoch at which the flux density of the radio flare reached its peak, is 
     shown for comparison. The two-point spectral indices between 843 MHz and 
     9.2 GHz are indicated at the bottom left of each panel. 
     Data are adopted from Hannikainen et al. (2000).}
\label{fig:spec}
\end{figure}

\begin{figure}
\setlength{\unitlength}{1in}
\begin{picture}(3.25,3.6)
\includegraphics{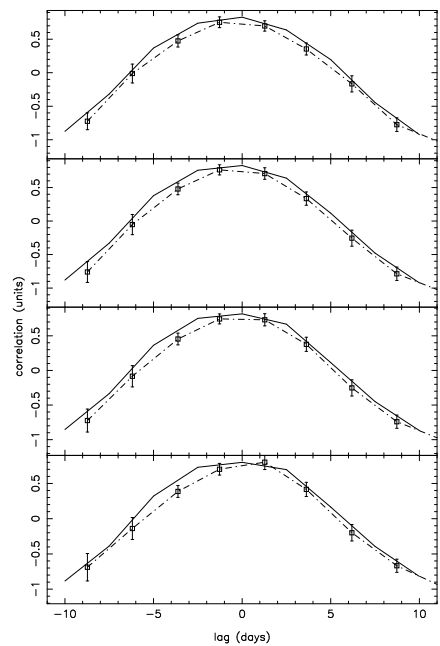}
\end{picture}
\caption{Application of the Discrete Correlation Function (points and
dot-dashed line) and Interpolated Correlation Function (solid line) to the
9.2 GHz versus (from top to bottom) 5.9, 2.3, 1.4 and 0.843 GHz data
streams.}
\label{fig:delay}
\end{figure}

The multifrequency flux density light curves and spectra (from Hannikainen et
al. 2000) are shown in Figs~\ref{fig:lc} and \ref{fig:spec} respectively. The
flux densities include the emission from the entire structure shown in the VLBA
images, with the possible exception of the two highest frequencies at which the
Australia Telescope Compact Array (ATCA) synthesized beam size is approximately
1 arcsec compared to a source size of 1--2 arcsec at 1.6 GHz. Our flux density
measurements involve fitting a point-source model to the data which may lead to
a slight systematic underestimate of the flux at 8.6 and 9.2 GHz. However, the
source may well be more compact at higher frequencies due to the shorter
synchrotron lifetimes of the emitting electrons (see Section~4), so it is
unlikely that much/any flux density is missing from our measurements. Indeed, in
Section~4, we argue that most of the flux density variability is confined to a
region $\ll 1$ arcsec in extent.  Since the available observations do not allow
separation of emission from the compact variable component(s) and the extended
regions we are forced to assume that the radio emission is dominated by a single
variable component.  The light curves show complex flaring behaviour with
several minor events following the initial increase in flux density. The key
features of the data are that the initial rise in flux density {\em peaks
simultaneously\/} at all frequencies and that the amplitude of the flare,
defined as maximum minus minimum flux density, {\em increases\/} towards lower
frequency (see Section~4 for further discussion). We quantified the time delays
between the data streams with the Discrete Correlation Function (DCF;  Edelson
\& Krolik 1988) and the Interpolated Correlation Function (ICF; Gaskell \&
Peterson 1987). Both methods give essentially the same result
(Fig.~\ref{fig:delay}); the correlation functions are approximately symmetrical
around zero lag in all cases. The flat peak to the curves suggest formal time
delays of $\sim 0\pm1$~d.

The first panel of Fig.~\ref{fig:spec} shows a spectrum from two days before
flux density maximum, which is consistent with synchrotron emission that is
optically thin at the highest frequencies but is partially self-absorbed at
around 1 GHz. As the flux density increases, the entire spectrum becomes
optically thin and remains so throughout the subsequent decline in flux
density.
In Fig.~\ref{fig:spec} we can see that the spectrum was relatively flat (with a
spectral index of $-0.41$ on TJD 9580). It steepened at the time of the maximum
(with the spectral indices reaching $-0.65$ on TJD 9582) and then flattened as
the flux density declined to a level of $-0.4$ around TJD 9586 and 9587. This
behaviour can be explained by models incorporating emission from multiple
regions.  In the simple core-lobe model suggested by Hannikainen et al. (2000),
the observed spectrum is the sum of the emission from a compact region (core),
which has a flat spectrum, and the emission from extended regions
(lobes/ejecta) with steeper spectra. The outburst can be seen as a consequence
of the brightening of the compact region and then the rapid rise in the
brightness of the ejecta. The emission from the ejecta eventually dominates as
the radio flux densities reach their peak amplitude. At the same time the
spectrum steepens. As the flare subsides, the emission from the ejecta drops to
a level comparable to, or lower than, the emission from the compact region, and
the spectra become flatter, similar to those seen at the onset of the flare.

The time-dependent properties of the radio outbursts of GRO~J1655$-$40 and
other black-hole X-ray binaries have been attributed to synchrotron emission
from expanding clouds of relativistic electrons. Under the usual assumption
that all radio frequencies are initially optically thick, the two major
characteristics of these models are that the low-frequency emission has a time
delay with respect to that at high frequencies and that the flare amplitudes
fall off towards lower frequency (see e.g.  Han \& Hjellming 1992).  The
observations clearly do not conform to these predictions of the expanding-cloud
model (see also Hannikainen et al. 2000).

If electrons are injected into an expanding cloud, and if expansion losses
dominate over radiative losses, then it is possible to get a simultaneous peak
at optically thin frequencies once the injection stops or once the expansion
cooling dominates over the injection. Note that if radiative losses dominate
under this scenario we would observe frequency-dependent peak times, assuming
the data streams are sufficiently well sampled.  However, the first panel of
Fig.~3 provides evidence that the synchrotron self-absorption turnover was close
to 1~GHz near the peak of the flare, implying that at least some of the radio
frequencies were optically thick at earlier epochs, and thus arguing against
this possibility. Similar scenarios can be produced by relaxing some of the
initial assumptions of the model. For example, Ball \& Vlassis (1993) considered
cases where electrons are injected into the cloud with a constant energy
spectrum. While both of these options can in principle reproduce the observed
behaviour (optical depth arguments aside) they require a specific set of
conditions that are unlikely to be met in practice.  Indeed, Marscher \& Gear
(1985) found that they could fit the variability of 3C~273 with a uniform
expanding source model but only if `the injection of relativistic electrons was
allowed to vary with radius in a rather ad hoc fashion'. They attributed the
success of this model to its large number of free parameters.  Arguing along
similar lines, we do not consider that such models provide a convincing physical
explanation for the phenomena observed in GRO~J1655$-$40. Instead, given the
obvious analogy between Galactic and extragalactic jet sources, we argue
alternatively that the variable radio emission arises in a small region behind a
relativistic shock wave, and that the rise of the flare is driven by
inverse-Compton cooling (cf. Marscher \& Gear 1985). We note that the shock
model developed by Kaiser et al. (2000) incorporates adiabatic and synchrotron
cooling only and as such is mostly applicable to the decline of radio
flares rather than to the rise phase discussed here. 

\section{Generalised shock model for extragalactic sources}  

\begin{figure}
\setlength{\unitlength}{1in}
\begin{picture}(3.25,2.6)
\includegraphics{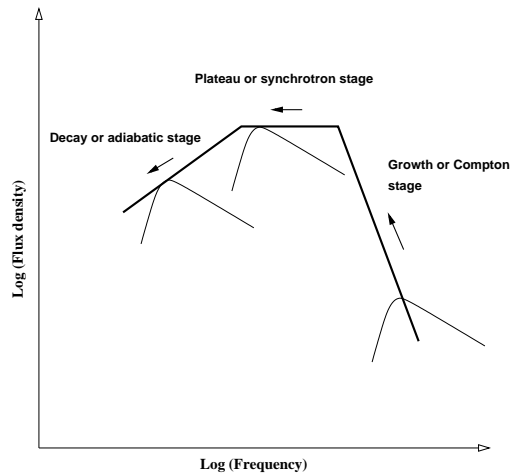}
\end{picture}
\caption{ Schematic of flare spectral evolution according to the shock model of
Marscher \& Gear (1985). The thick solid line and arrows show the time evolution of the
synchrotron self-absorption turn-over ($S_{\rm m}$) as the shock expands along
the jet.}
\label{fig:mg}
\end{figure}

Spectral evolution during the flares of compact extragalactic radio sources can
be explained by the generalised shock model of Marscher \& Gear (1985). (For
specific applications see e.g. Valtaoja et al. 1988 and Stevens et al. 1994,
1996, 1998).
In the model, temporal evolution of the radio emission is divided into growth,
plateau and decay stages (see the schematic illustration in
Fig.~\ref{fig:mg}). The emission region consists of relativistic jets along
which transverse shocks propagate and accelerate electrons to relativistic
energies.  The shock waves form in response to changes in the conditions in the
jet (such as pressure or velocities of the bulk flow). The evolution of the
shock is described in terms of the flux density ($S_{\nu}$) at the peak
frequency ($\nu_{\rm m}$ where $S_{\nu}[\nu_{\rm m}]=S_{\rm m}$) of the
synchrotron spectrum where the opacity is close to unity. This point is fixed
at any one time by the dominant energy-loss mechanism.  When the emitting
region is compact, inverse-Compton losses predominate (Compton or growth stage)
but these fall off rapidly with radius as the shock expands and are superseded
by synchrotron losses (synchrotron or plateau stage). As the shock expands
further, the radiative lifetime of the electrons becomes large with respect to
the time needed to traverse the emitting region and losses due to adiabatic
expansion (adiabatic or decay stage) become more important.  All three stages
are approximated by power laws on the logarithmic $(S_{\rm{m}},\nu_{\rm{m}}$)
plane. As the emitting region expands, $\nu_{\rm m}$ is predicted to move to
lower frequencies with time.  $S_{\rm{m}}$ increases rapidly during the Compton
stage, remains approximately constant during the synchrotron stage and
decreases during the adiabatic stage.

The characteristics of spectral evolution predicted by the generalised shock
model are summarised as follows (see Valtaoja et al. 1992 for more details).
The maximum flare amplitude for any frequency on the Compton stage
occurs when the spectrum transits onto the synchrotron stage. Light curves at
such frequencies are thus predicted to peak
simultaneously and because of the spectral shape, the flare amplitudes are
expected to increase towards lower frequency -- specifically, they should have
the same power law form as the optically thin portion of the flare spectrum.
Flare amplitudes at frequencies commensurate with the later stages
are determined by the details of the spectral evolution, and are thus expected
to be approximately constant during the synchrotron stage and to decrease
during the adiabatic stage. The light curves will display time-lagged behaviour
with a delay between any two frequencies equal to the time taken for the
spectrum to evolve between them and become optically thin. Note that the
development of the flare during the adiabatic stage is very similar to that
predicted by the expanding cloud model, although, because the expansion is
constrained to occur in a jet, the predicted time delays are shorter and the
variation of flare amplitude with frequency is less pronounced.
   
\section{Microquasars vs quasars}  

\begin{figure}
\setlength{\unitlength}{1in}
\begin{picture}(3.25,2.0)
\includegraphics{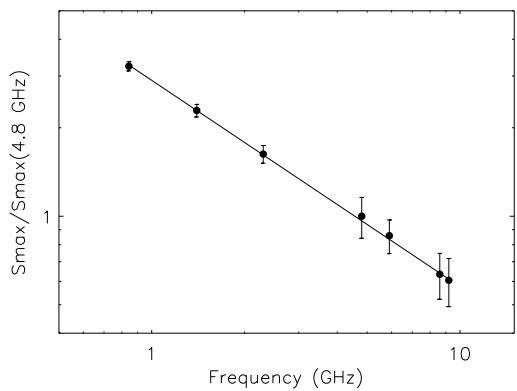}
\end{picture}
\caption{Flare amplitudes relative to 4.8 GHz. The slope of the fitted line
is $-0.70\pm0.05$}
\label{fig:amp}
\end{figure}

The generalised shock model prediction of simultaneously peaking emission at
all frequencies coupled with flare amplitudes that increase towards lower
frequencies (i.e. the Compton/growth stage) is consistent with the observations
of GRO~J1655$-$40 in 1994.  According to the shock model, we would also expect
the flare amplitudes to fall off as a power law since they should have the same
frequency dependence as the spectrum of the flare when it reaches the
transition point between the growth and plateau stages. In Fig.~\ref{fig:amp}
we plot the flare amplitudes normalised to 4.8 GHz. A power law is clearly a
good fit with a slope of $-0.70\pm0.05$, consistent with that expected for the
optically thin portion of the flare synchrotron spectrum (cf. Stevens et
al. 1996).

One apparent difference between the flaring behaviour of GRO~J1655$-$40 and its
extragalactic analogues, the blazars, is that the flare remains in the growth
stage to much lower frequencies. For example, many of the sources in the sample
considered by Stevens et al. (1994) show time delays between the 37 GHz, or
occasionally the 90 GHz, emission and that at higher frequencies, although
those authors noted that some BL~Lacertae objects remained in the growth stage
out to 4.8 GHz. An obvious contributor to this difference is the Doppler effect
which will shift the emitted frequency of radiation in the jet frame, $\nu'$,
to a higher observed frequency, $\nu = \delta\nu'$.  The relativistic Doppler
factor, $\delta=\Gamma^{-1}(1-\beta \ {\rm cos} \ \theta)^{-1}$ where
$\Gamma=(1-\beta^2)^{-1/2}$ is the bulk Lorentz factor, $\theta$ is the viewing
angle and $\beta$ is the jet speed in units of $c$. It is generally accepted
that blazars have their jet axes oriented quite closely towards the observer;
for example Ter\"{a}sranta \& Valtaoja (1994) find that blazars on average are
aligned to within about 20 deg of the line-of-sight (see also Barthel et
al. 1989). Estimates of Doppler factors for blazars range from extreme values
of 0.005 to 33 with a mean value of around 5 (Guijosa \& Daly 1996). The
blazars with measured time delays from the sample of Stevens et al. (1994) have
`inverse-Compton' Doppler factors in the range 3.4--16 (see Guijosa \& Daly
1996).  The microquasar GRO~J1655$-$40, however, is angled further towards the
plane of the sky ($\theta=70$ deg; Orosz \& Bailyn 1997) and its estimated jet
speed $\beta=0.92$ (Hjellming \& Rupen 1995) leads to $\delta\sim0.6$. Higher
values of the Doppler factor for blazars compared to that of GRO~J1655$-$40
could thus easily lead to an order of magnitude difference in the frequency at
which the flare exits the growth stage.

A second contributory factor might arise from physical differences
between the jets of Galactic and extragalactic sources. Since 
the flare evolution is dependent on the photon energy density, the magnetic
energy density and the lifetime of the emitting electrons compared to the time
they take to cross the shock structure, two important parameters are the
size of the emitting region and its magnetic field strength. 

We can estimate the magnetic field from the flare decay time by assuming that
the decay is radiative rather than expansion-loss driven.  For the clear-cut
case of a single flare we would expect the 1/e decay times, $t$, of the light
curves to vary as $\nu^{-1/2}$ if radiative losses are the dominant energy-loss
mechanism, or if adiabatic expansion losses predominate, then $t$ will be the
same at all frequencies. Unfortunately, the flaring behaviour that we observe
is complex. At the monitoring frequencies presented in Fig.~\ref{fig:lc} the
decay of the flares is interrupted by subsequent events, making it impossible
to measure an accurate e-folding timescale. However, the multifrequency light
curves presented by Hjellming \& Rupen (1995) include data up to 22.5 GHz. For
the flare in question, the decay of the 22.5 GHz lightcurve is more rapid than
those at lower frequencies, most probably because the monitoring frequency is
now sufficiently high that the e-fold timescale is similar to the interval
between the bursts. This result provides evidence that synchrotron losses are
driving the flux decay. If this is not the case then the implication would be
that the synchrotron-loss stage is short lived or non-existent, as observed for
3C\ 273 (Stevens et al. 1998). The rest of this section assumes that the decay
of the flare is driven by synchrotron losses. If this is not the case, and
adiabatic expansion losses predominate then we will overestimate the magnetic
field strength and underestimate the electron Lorentz factor and source extent.

For an e-fold decay timescale $t$ and a peak frequency $\nu_{\rm m}$, the
magnetic field is given by
\begin{equation}
B \sim 23\,\delta^{-1/3} \left ( \frac{\nu_{\rm m}}{1~{\rm GHz}} \right )^{-1/3}
\left ( \frac{t}{5~{\rm day}} \right )^{-2/3}~{\rm gauss} \ . 
\end{equation} 
We estimated the flare decay time from the best sampled light curve
(843 MHz). For the reasons discussed above this estimate is likely to be an
upper limit but should not be out by more than a factor of two; using 5 d
the estimated magnetic field strength is $\sim$25~gauss. This value is
approximately two orders of magnitude higher than typical estimates of 0.1--1
Gauss in extragalactic sources (e.g. Brown et al. 1989).

The Lorentz factor of the electrons emitting at 843 MHz is given by
$\gamma\sim(2 \pi m_{\rm e} c \nu/eB)^{1/2}\sim3.5$, where $e$ and $m_{\rm e}$
are the electron charge and mass respectively.  This value is approximately
that required to explain the observed circular polarization as being intrinsic
to the synchrotron radiation (Macquart et al. 2002) although the alternative
mechanism, Faraday conversion, could also apply.  In any case, we note that the
presence of circularly polarized radiation is another characteristic that
GRO~J1655$-$40 shares with the blazars (see e.g. Wardle et al. 1998).
Furthermore, strong evidence for our proposed model comes from the observed
position angle of linear polarization at radio wavelengths which is very close
to that of the jet direction (Hannikainen et al. 2000). For optically thin
synchrotron radiation the implied magnetic field direction is perpendicular to
the jet, as observed for many blazars (e.g. Cawthorne et al. 1993a,b), and as
predicted by models in which a tangled component of magnetic field in the jet
is compressed parallel to a transverse shock front (e.g. Hughes et al. 1989a).
 
Finally, we can estimate the source extent at the time when the spectrum
transits from the growth stage onto the plateau stage. At this transition, the
energy densities in particles ($U_{\rm ph}$) and magnetic field are equal and
thus so are the synchrotron and inverse-Compton lifetimes $t_{\rm syn}=t_{\rm
IC}$. Using
\begin{equation} 
t_{\rm IC} \sim {{3\times 10^7} \over {\gamma U_{\rm ph}}} \sim 
{{3\times 10^7 cD^2} \over {\gamma \phi L}} \ {\rm seconds} \ ,
\end{equation} 
and assuming that, (1) the transition occurs when the 843 MHz light curve peaks
(i.e. the flare amplitudes peak at this frequency), allowing us to use the
values $t_{\rm IC}=t_{\rm syn}\sim 5$~d and $\gamma \sim 3.5$ deduced above,
(2) the radiation field is produced by a central source of luminosity,
$L\sim10^{38}$ erg\,s$^{-1}$ and (3) that $\phi$, a geometrical constant, is
approximately unity, the calculated source size $D \sim 10^{13}$~cm. At a
source distance of 3.2 kpc (Hjellming \& Rupen 1995) this dimension translates
to an angular size of about 0.2 mas. A 1.6 GHz VLBA map taken one day after the
peak of the 843 MHz light curve shows a source of about 1 arcsec with a bright
core and extended jets (Hjellming \& Rupen 1995).  Our proposed model requires
that the radio flares originate in a small region close to the base of the jet
as is observed in AGN.

\section{Final remarks} 

We have shown that a model incorporating a propagating relativistic shock in a
jet can qualitatively reproduce the multifrequency variability of
GRO~J1655$-$40. This model, which also fits the available polarization
observations, is the canonical model used to explain the variability
characteristics of extragalactic radio sources such as blazars.  A detailed
discussion of the applicability of this model to Galactic superluminal sources
is beyond the scope of the present work, and in any case is limited by the
availability of suitable data sets. It is important that other flares similar
to the one discussed here are observed over a broad range of radio frequencies,
and that they are first observed on the rising portion of the light curves. In
this respect, we point out that the two sources discussed by Ball \& Vlassis
(1993), Nova Muscae 1991 and the Galactic Centre Transient, for which data were
taken during the rise of the flares, both display the behaviour predicted by
the Compton stage of the shock model. Both sources were observed to have an
optically thin spectrum during the rising phase of the outburst which can be
accommodated by the Compton stage if the self-absorption turnover had already
passed the lowest observing frequency when the flux monitoring began.
  
\section*{Acknowledgments}  

J.A.S. and D.C.H. acknowledge support from PPARC, and K.W. acknowledges support
from the Australian Research Council through an Australian Research
Fellowship. D.C.H also acknowledges travel support from the Academy of Finland,
and thanks the University of Sydney and Jodrell Bank for their hospitality and
financial support during her visits.

\end{document}